\documentclass[9pt,shortpaper,twoside,web]{ieeecolor}
\usepackage{generic}
\usepackage{cite}
\usepackage{amsmath,amssymb,amsfonts}
\usepackage{algorithmic}
\usepackage{graphicx}
\usepackage{textcomp}
\usepackage{url}

\usepackage{amsthm}
\usepackage{cancel}
\def\BibTeX{{\rm B\kern-.05em{\sc i\kern-.025em b}\kern-.08em
    T\kern-.1667em\lower.7ex\hbox{E}\kern-.125emX}}
\theoremstyle{plain}
\newtheorem{theorem}{Theorem}
\newtheorem{lemma}{Lemma}

\newtheorem{proposition}{Proposition}

\theoremstyle{assumption}
\newtheorem{definition}{Definition}

\newtheorem{remark}{Remark}
\theoremstyle{goal}
\newtheorem{goal}{Goal}
\theoremstyle{assumption}
\newtheorem{assumption}{Assumption}

\begin{document}
\title{Regression Filtration with Resetting to Provide Exponential Convergence of MRAC for Plants with Jump Change of Unknown Parameters}
\author{Anton Glushchenko, \IEEEmembership{Member, IEEE}, Vladislav Petrov, and Konstantin Lastochkin
\thanks{Research was partly financially supported by Grants Council of the President of the Russian Federation (project MD-1787.2022.4).}
\thanks{A. I. Glushchenko is with V.A. Trapeznikov Institute of Control Sciences RAS, Moscow, Russia (phone: +79102266946; e-mail: aiglush@ipu.ru).}
\thanks{V. A. Petrov is with Stary Oskol technological institute (branch) NUST “MISIS”, Stary Oskol, Russia (e-mail: petrov.va@misis.ru).}
\thanks{K. A. Lastochkin is with V.A. Trapeznikov Institute of Control Sciences RAS, Moscow, Russia (e-mail: lastconst@ipu.ru).}}

\maketitle

\begin{abstract}
This paper proposes a new method to provide the exponential convergence of both the parameter and tracking errors of the composite adaptive control system without the persistent excitation (PE) requirement. Instead, the derived composite adaptive law ensures the above-mentioned properties under the strictly weaker finite excitation (FE) condition. Unlike known solutions, in addition to the PE requirement relaxation, it provides better transient response under jump change of the plant uncertainty parameters. To derive such an adaptive law, {a novel scheme of uncertainty filtration with resetting is proposed, which provides the required properties of the control system.} A rigorous proof of all mentioned properties of the developed adaptive law is presented. Such law is compared with the known composite ones, which also relax the PE requirement, using the wing-rock problem to conduct numerical experiments. The obtained results fully support the theoretical analysis and demonstrate the advantages of the proposed method.
\end{abstract}


\section{Introduction}
\label{sec:introduction}

Several groups of adaptive observation and control methods have been developed specifically to control the plants with significant parameter uncertainty \cite{b1}. The first one includes adaptive observers and adaptive laws to estimate the parameters of the plant or the disturbance ({\it Model Reference Adaptive Systems} – MRAS) \cite{b1,b2,b3}. The second group consists of methods of direct, indirect, and composite { \it Model Reference Adaptive Control} (MRAC) \cite{b4,b5}. Both groups require to identify the unknown parameters of the plant uncertainty, which, in their turn, are used to estimate the plant states (MRAS) or compensate for the influence of the uncertainty on the control quality (MRAC). To achieve this, the plant uncertainty is expressed in the linear regression form. Then the second Lyapunov method is applied to derive the adaptive law to estimate its unknown parameters \cite{b6}.

A well-known and deeply investigated drawback of the Lyapunov-based identification is that the estimates converge exponentially to the ground-truth values only when the requirement of the regressor {\it persistent excitation} (PE) is met \cite{b6,b7,b8}. Generally speaking, the parameter convergence is a very advantageous property considering MRAC and MRAS \cite{b8}, because it automatically guarantees exponential convergence of the tracking error (between the states of the observer and the plant for MRAS or the states of the plant and the reference model for MRAC) \cite{b6}. It also provides the robustness of the adaptive law to exogenous bounded disturbance. If the PE condition is not satisfied, such a law requires the application of some robust modifications \cite{b6,b9}. It was proved in \cite{b10} that the PE requirement is satisfied if the number of the spectral lines in the reference signal coincides with the number of the regression unknown parameters. For many practical applications, the fulfillment of this condition (as a result of the reference signal modification) could be inconsistent with the initial control objective, and also leads to increased power consumption and wear of the actuators. So, in recent years the aim to provide exponential parameter convergence under conditions, which are strictly weaker than PE, has become of high interest. We recommend \cite{b11} and references therein as a good survey of most of the existing solutions to that problem. While only main methods to relax PE in MRAC are considered in the following analysis.

Conventional adaptive law requires PE condition to provide the exponential convergence as the objective of such law is minimization of the function of the instantaneous (proportional) tracking error. In case when the PE condition is not met, such function may have a minimum at points, which do not coincide with the one corresponding to the regression ideal parameters \cite{b12}. Therefore, the main concept of most of the known methods of PE requirement relaxation is the transformation of the problem from the optimization of the proportional objective function to the proportional-integral one, because it has a single minimum at the point corresponding to the ideal parameters even in the {\it finite excitation} (FE) case \cite{b12}.

Considering MRAC, this concept can be implemented using the ideas of the {\it composite adaptation method} (CMRAC) \cite{b5}. According to it, the adaptive law includes two summands: 1) to minimize the tracking error as the difference between the states of the plant and the reference model, 2) to estimate the ideal parameters of the plant uncertainty. Further modifications of CMRAC are aimed at optimization of the integral error of the uncertainty identification. For this purpose, BackGround \cite{b13,b14,b15}, {\it Concurrent Learning} (CL) \cite{b16,b17}, and PI adaptive law \cite{b18} have been proposed to obtain an integral uncertainty identification error and thereby relax the PE requirement to the finite (FE) or initial (IE) excitation ones.

To obtain the integral error, it is proposed \cite{b13,b14,b15,b16,b17} to save the data on the uncertainty in the DataStack. For the same purpose, it is proposed \cite{b18} to use an open loop integration in the {\it memory regression extension} (MRE) procedure \cite{b11,b19} instead of the Kreisselmeier filter \cite{b19}. The main common problems of these two methods are, firstly, unbounded growth of the information matrix (integral regressor) in the case of noise and disturbances or when the PE condition holds. And, secondly, inaccurate identification of the uncertainty piecewise-constant parameters. To solve the first problem, it is proposed in \cite{b20} to use, considering MRE, not a purely integral filter, but the one with exponential forgetting. Unlike \cite{b13,b14,b15,b16,b17,b18}, it always allows to obtain a bounded regressor. But this method, as well \linebreak as \cite{b13,b14,b15,b16,b17,b18}, is not able to identify the switched parameters correctly.

Inaccurate identification of the piecewise-constant parameters is caused by the fact that in this case the integral objective function includes information not about one uncertainty (regression), but about some superposition of several of them. Consequently, the minimum of such objective function is at the point, which is the averaged value of the ideal parameters of all accumulated regressions \cite{b12}. Such superposition occurs because the methods \cite{b13,b14,b15,b16,b17,b18,b20} do not have an algorithm to forget completely the already used and {\it outdated data} about uncertainty. As a solution to this problem, let the methods be considered, which relax the PE requirement and simultaneously have some forgetting property for outdated data.

First of all, it is the approach \cite{b21} that use some production rule to switch from integral filtering \cite{b18} to aperiodic Kreisselmeier’s one \cite{b19}. The disadvantage of this method is that the outdated data forgetting procedure is executed only when a special condition on the product of the parameter error and the value of the unknown parameters change holds. This significantly reduces the applicability domain of such method. Secondly, it is \cite{b22}, in which it has been proposed to use a variable forgetting factor in the Kreisselmeier filter \cite{b19} and a regression to identify the plant parameters, which is obtained when some filtered regressor metric has maximum value. Such a metric is calculated on the finite excitation intervals. Since this approach does not provide strict guarantees on the improvement of the maximum value of the chosen metric, then the update of regressions, which are used for the parameter estimation, may not happen. So, this method provides exponential parameter error convergence only if the plant unknown parameters are time-invariant. {Thirdly, in \cite{b22i5} it is proposed to change the filtering forgetting factor \cite{b19} proportionally to the value of the minimum eigenvalue of the filtered regressor. Particularly, it is increased when the regressor excitation level is high, and decreased to zero otherwise. Zero value of the forgetting factor allows one to ensure that the filtered regressor does not vanish and, therefore, relax the PE requirement for the parameter error exponential convergence. In turn, high value of the forgetting factor improves the rate of forgetting of the outdated data on the uncertainty compared to filtering with a time-invariant forgetting factor \cite{b19}. However, this approach guarantees that the outdated data on the uncertainty are completely forgotten only when PE is met, and requires the minimum and maximum values of the minimum eigenvalue of the filtered regressor to be known a priori to effectively adjust the filter parameter \cite{b19} and improve the forgetting quality.} Thus, considering the MRAC schemes, none of the above-mentioned approaches ensures the exponential convergence of the estimation process of the plant piecewise-constant parameters without PE.

Therefore, in this research, a new method is proposed, which guarantees exponential convergence of the estimation process of the plant unknown piecewise-constant parameters in case the regressor is FE. {It is proposed to reset the output of all filters, which are used in the adaptive control system parametrization, to zero instantaneously at time points, which corresponds to a piecewise-constant reference signal $r\left(t\right)$ value change.} The need for instantaneous resetting of all filters, rather than aperiodic forgetting, is motivated by \cite{b21, b22, b22i5}, according to which, when aperiodic forgetting is applied, often the filter output does not have enough time to reach zero (or sufficiently low value) in the absence of PE. This leads to inaccurate identification of the unknown switched parameters under FE condition.

A similar idea has been applied to implement Concurrent Learning Model Predictive Control system \cite{b23}. In this approach, when some kind of plant identification error metric is low, the model is known with high accuracy and MPC is applied. If such metric becomes high, i.e. the ideal parameters of the plant have changed and need to be identified again, the data stack of CL is cleared, and CMRAC is used again. But this approach is not able to track piecewise-constant unknown parameters without switching to MPC stage. The algorithm may be stuck at CMRAC stage in case the ideal values of unknown parameters change in the course of it. Therefore, unlike \cite{b23}, we propose to apply a resetting procedure to relax PE requirement for MRAC, using reference signal $r\left(t\right)$ as an indicator to reset the filter.

The novel features and main contributions of this paper are summarized as follows: {1) the application of a resetting procedure to update dynamically the outputs of all filters, which are used in the adaptive control system parametrization;} 2) the PE requirement is relaxed to FE one to provide the exponential convergence of the plant unknown piecewise-constant parameters estimates for MRAC.

{It is proposed to combine and improve a number of recent results \cite{b18,b19,b20,b25} in order to achieve the stated goals and main contributions. (i) Introducing a resetting procedure into a method of uncertainty parameterization on the basis of the aperiodic filtering [18], it is proposed to obtain a measurable value of the filtered uncertainty in a linear regression form. (ii) Using Kreisselmeyer filtering with resetting as a part of the \emph{dynamic regressor extension and mixing} (DREM) procedure \cite{b25}, it is proposed to transform the matrix regressor obtained at the first step into a scalar one to improve the identification quality. (iii) To obtain a composite adaptive law with exponential convergence of the parameter error under FE condition, it is proposed to augment the integral filter with exponential forgetting [20] with a resetting procedure. Such procedure for the filters of (i)-(iii) is proposed to be executed when the reference value is changed, which makes it possible to completely and instantaneously exclude the influence of outdated data on uncertainty on the control quality.}

The rest of the paper is organized as follows. Section II presents the notation used in the paper; Section III gives a generalized problem statement; Section IV presents the proposed filtering procedure and adaptive law; in Section V the exponential convergence of the proposed CMRAC is proved; Section VI compares it with the known methods; Section VII presents the results of simulation.

\section{Preliminaries}

The following notations and definitions are used in the paper: ${L_\infty }$ is the space of the essentially bounded functions, {${L_2}$ is the space of quadratically integrable functions}, ${\lambda _{\min }}\left( . \right)$ and ${\lambda _{\max }}\left( . \right)$ are the minimal and maximal eigenvalues of a matrix, $vec\left( . \right)$ is the operation of a matrix vectorization, $\left\| . \right\|$ is the Euclidean norm of a vector, ${\left\| . \right\|_{\rm{F}}}$ is the Frobenius norm of a matrix, {$o\left( . \right)$ means “is ultimately smaller than”, $tr(.)$ and $(.)^{{\rm{\dag }}}$ are the matrix trace and the Moore-Penrose pseudo-inverse operators, $det \left\{ . \right\}$ stands for a matrix determinant, $adj\left\{ . \right\}$ – for an adjoint matrix, ${I_{n \times n}}$ is an identity $n \times n$ matrix, ${0_{n \times n}}$ is a zero $n \times n$ matrix, $exp\left(.\right)$ stands for an exponential function,} $f\left( t \right)$ is a function, which depends on time (the time argument is omitted when it does not cause any confusion).

\begin{definition}\label{definition1}
A regressor $\varphi \left( t \right) \in {L_\infty }$ is finitely exciting $\left( {\varphi \left( t \right) \in {\rm{FE}}} \right)$ over a time set $\left[ {{t_s}{\rm{;}\;}{t_s} + T} \right] \subset \left[ {{t_0}{\rm{;}}\;\infty } \right]$ if there exist ${t_s} \ge {t_0} \ge 0$, $T > 0$  and a level of excitation $\alpha  > 0$ such that
\begin{equation}\label{eq1}
	\int\limits_{{t_s}}^{{t_s} + T} {\varphi \left( \tau  \right){\varphi ^{\rm{T}}}\left( \tau  \right)d} \tau  \ge \alpha {I_{n \times n}}.
\end{equation}
\end{definition}

Let $\dot x\left( t \right) = f\left( {x\left( t \right)} \right)$ be a dynamic system with stable origin and globally Lipshitz continuous right-hand side. Then:

\begin{definition}\label{definition2}
A system equilibrium $\forall t \ge {t_0} + T$ is globally exponentially stable $\left( {x\left( t \right) \in {\rm{GES}}} \right)$ if there exists $\kappa>0$ and $\rho>0$, such that $\left\| {x\left( t \right)} \right\| \le \rho \left\| {x\left( {{t_0} + T} \right)} \right\|{e^{ - \kappa \left( {t - {t_0} - T} \right)}}$ for any $x\left( {{t_0} + T} \right)$.
\end{definition}

\begin{definition}\label{definition3}
The solution $x\left( t \right)$ is exponentially ultimately bounded $\left( {x\left( t \right) \in {\rm{EUB}}} \right)$ with uniform ultimate bound $R > 0$ if for $\kappa  > 0$, $\rho  > 0$ there exists a time instant ${t_0} + T$, such that $\left\| x \right\| \le \linebreak \!\le\!\! \rho \left\| {x\left( {{t_0} \!+ \!T} \right)} \right\|\!{e^{\! -\! \kappa \left( {t - {t_0} - T} \right)}}\! +\! R$ for $t \!\ge\! {t_0}\! +\! T$ and any state $x\!\left( {{t_0}\! +\! T} \!\right)$.
\end{definition}

\section{Problem Statement}

\subsection{{Systems Dynamics}}

The class of {\it linear time-invariant} (LTI) plants is considered:
\begin{equation}\label{eq2}
\forall t \ge {t_0}{\rm{}\;}\dot x\left( t \right) = Ax\left( t \right) + B\left( {u\left( t \right) + \Delta \left( t \right)} \right){\rm{,}\;}{x\left( {{t_0}} \right) = {x_0}{\rm{,}}}
\end{equation}
where $x\left( t \right) \in {\mathbb{R}^n}$ is a state vector with known initial conditions ${x_0}$, $u\left( t \right) \in {\mathbb{R}^m}$ is a control action, $\Delta \left( t \right) \in {\mathbb{R}^m}$ is a parameter uncertainty, $A \in {\mathbb{R}^{n \times n}}$ is a known state matrix, and $B \in {\mathbb{R}^{n \times m}}$ is a known input matrix of full column rank. Pair $\left( {A{\rm{,}}\;B} \right)$ is controllable such that $m \le n$. The vectors $x\left( t \right){\rm{,}}\;u\left( t \right)$ are measurable.

The uncertainty $\Delta \left( t \right)$ can be linearly parametrized such that:
\begin{equation}\label{eq3}
\Delta \left( t \right) = {\Theta ^{\rm{T}}}\left( t \right)\Phi \left( {x\left( t \right)} \right),
\end{equation}
where $\Phi \left( x \right) \in {\mathbb{R}^p}$ is a bounded measurable regressor, $\Theta \left( t \right) \in {\mathbb{R}^{p \times m}}$ is a matrix of unknown parameters, for which the following holds.

\begin{assumption}\label{assumption1}
{The parameters $\Theta \left( t \right){\rm{}\;\!}\forall t \ge {t_0}$ have jump behavior:}
\begin{equation}\label{eq4}
{\dot \Theta \left( t \right) = \sum\limits_j {{\theta _j}\delta \left( {t - {t_j}} \right)} {\rm{,}\;}}\Theta \left( t \right) = {\Theta _0} + \sum\limits_j {{\theta _j}h\left( {t - {t_j}} \right)} {\rm{,}}
\end{equation}
where ${\theta _j} \in {\mathbb{R}^{p \times m}}$ is the value of the unknown parameter change, $h\left( t \right)$ is the step function, {$\delta \left( t \right)$ is the delta function,} ${t_j} > {t_0}$ is the unknown time instant when the unknown parameters change.
\end{assumption}

So, the problem of the adaptive control of the LTI plant \eqref{eq2} with the piecewise-constant parameters of uncertainty $\Delta \left( t \right)$ is considered. 

\subsection{{Reference Model Dynamics}}

The reference model for plant \eqref{eq2} is chosen as:
\begin{equation}\label{eq5}
{\dot x_{ref}}\left( t \right) = {A_{ref}}{x_{ref}}\left( t \right) + {B_{ref}}r\left( t \right){\rm{,}\;}{{x_{ref}}\left( {{t_0}} \right) = {x_{0ref}}{\rm{,}}}
\end{equation}
where ${x_{ref}}\left( t \right) \in {\mathbb{R}^n}$ is the reference model state vector, $r\left( t \right) \in {\mathbb{R}^m}$ is the reference signal, ${A_{ref}} \in {\mathbb{R}^{n \times n}}{\rm{,}}\;{B_{ref}} \in {\mathbb{R}^{n \times m}}$. The matrix ${A_{ref}}$ is chosen so as the equation $A_{ref}^{\rm{T}}P + P{A_{ref}} =  - Q$ has a solution $P$ with $Q = {Q^{\rm{T}}} > 0,{\rm{}}\;P = {P^{\rm{T}}} > 0$. Reference signal $r\left( t \right)$ for plant (2) is chosen so as to satisfy the following assumptions.
\begin{assumption}\label{assumption2}
The reference $r\left( t \right)$ is a piecewise-constant function:
\begin{equation}\label{eq6}
r\left( t \right) = \sum\limits_k {{r_k}h\left( {t - {t_k}} \right)},
\end{equation}
where ${r_k} \in {\mathbb{R}^m}$ is the value of the reference change, ${t_k} \ge {t_0}$ is a known time instant of the reference change.
\end{assumption}

\begin{assumption}\label{assumption3}
After each change of the reference value it holds that the regressor $\Phi \left( x \right) \in {\rm{FE}}$ over a time range $\left[ {{t_k}{\rm{;}}\;{t_k} + T} \right]$.
\end{assumption}

{The most preferable relationship between the time instants ${t_j}$ and ${t_k}$ for practical scenarios is shown in Fig. 1 of \cite{b24}.}

\subsection{{Control Law}}

The control law $u\left( t \right)$ to provide the control quality \eqref{eq5} to plant \eqref{eq2} is chosen as state feedback with direct uncertainty compensation:
\begin{equation}\label{eq7}
{\small
u\!\left( t \right)\!=\!{u_{bl}}\!\left( t \right)\!-\!{u_{ad}}\!\left( t \right)\!=\!{K_x}x\!\left( t \right)\!+\!{K_r}r\!\left( t \right)\!-\!{\hat \Theta ^{\rm{T}}}\!\left( t \right)\Phi\! \left( {x\left( t \right)} \right){\rm{,}}
}
\end{equation}
where ${K_x} \in {\mathbb{R}^{m \times n}}$, ${K_r} \in {\mathbb{R}^{m \times m}}$ are the parameters of the baseline part ${u_{bl}}\left( t \right)$, $\hat \Theta \left( t \right) \in {\mathbb{R}^{p \times m}}$ is the unknown parameters estimate.

The control law \eqref{eq7} is substituted into \eqref{eq2} to obtain:
\begin{equation}\label{eq8}
{\small
\dot x\!\left( t \right)\!=\!\left( {A\!+\!B{K_x}} \right)x\!\left( t \right)\!+\! B{K_r}r\left( t \right)\!+\!B\!\underbrace {\left( {{\Theta ^{\rm{T}}}\!\!-\!\!{{\hat \Theta }^{\rm{T}}}\left( t \right)} \right)}_{{{\tilde \Theta }^{\rm{T}}}\left( t \right)}\Phi \left( {x\left( t \right)} \right){\rm{,}}
}
\end{equation}
where $\tilde \Theta \left( t \right) \in {\mathbb{R}^{p \times m}}$ is the error between $\Theta \left( t \right)$ and $\hat \Theta \left( t \right)$. 

\subsection{{Tracking Error Dynamics}}
{The control objective is to make \eqref{eq8} behave as the reference model \eqref{eq5}. To do this, we assume that the parameters of the baseline part of \eqref{eq7} satisfy the following conventional matching conditions.}
\begin{assumption}\label{assumption4}
There exists ${K_x} \in {\mathbb{R}^{m \times n}}$, ${K_r} \in {\mathbb{R}^{m \times m}}$ such that:
\begin{equation}\label{eq9}
A + B{K_x} = {A_{ref}}{\rm{;}}\;B{K_r} = {B_{ref}}.
\end{equation}
\end{assumption}

If Assumption \ref{assumption2} is met, then the error equation between \eqref{eq8} and \eqref{eq5} is written as:
\begin{equation}\label{eq10}
{\small
\dot e_{ref}}\left( t \right) = {A_{ref}}{e_{ref}}\left( t \right) + B{\tilde \Theta ^{\rm T}}\left( t \right)\Phi \left( {x\left( t \right)} \right){\rm{,}}\;{{e_{ref}}\left( {{t_0}} \right)\!=\!{e_{0}}}{\rm{.}
}
\end{equation}

The augmented error ${{\xi \left( t \right) = { {\begin{bmatrix} {e_{ref}^{\rm{T}}\left( t \right)}&{vec\left( {{{\tilde \Theta }^{\rm{T}}}\left( t \right)} \right)} \end{bmatrix}} ^{\rm{T}}}}}$ is introduced into \eqref{eq10}. Then, we are in position to formulate the adaptive control goal for the plant \eqref{eq2}.

\begin{goal}
	Let the assumption \ref{assumption1}-\ref{assumption4} be met, then the task is to derive a law to adjust the parameters of the adaptive part ${u_{ad}}\left( t \right)$ such that the augmented error is globally exponentially stable $\left( {\xi \left( t \right) \in {\rm{GES}}} \right)$.
\end{goal}

\begin{remark}\label{remark1}
Assumption \ref{assumption2} is met for most real plants, excluding the ones from the inner loop of cascade control systems. The meaning of Assumption \ref{assumption3} is that each change of the reference value $r\left( t \right)$ in accordance with \eqref{eq6} results in the finite excitation of the $\Phi \left( x \right)$. Assumption \ref{assumption4} is a standard assumption for MRAC. Practically speaking, the value of ${t_k}$ can be obtained online using $r\left( t \right) \ne r\left( {t - {T_d}} \right) \Leftrightarrow t = {t_k}$, where ${T_d} > 0$ is sufficiently close to zero.
\end{remark}

\section{Main Result}

{In this section, a composite adaptive law is to be derived to meet the stated Goal. To obtain such a law: \textbf{in subsection A} the filtered value ${\Delta _f}\left( t \right) = {\Theta ^{\rm{T}}}\left( t \right){\Phi _f}\left( {x\left( t \right)} \right) \in {\mathbb{R}^m}$ of uncertainty \eqref{eq3} is expressed from \eqref{eq10}, \textbf{in subsection B}, applying DREM procedure, the matrix regressor ${\Phi _f}\left( x \right)$ is transformed into a scalar one $\omega \left( t \right) \in \mathbb{R}$, \textbf{in subsection C} the integral filtration with resetting and forgetting is proposed to obtain the regressor $\Omega \left( t \right) \in \mathbb{R}$, which does not vanish if $\Phi \left( x \right) \in {\rm{FE}}$, from $\omega \left( t \right) \in \mathbb{R}$ and derive a composite adaptive law with relaxed excitation requirements.}

\subsection{Plant Uncertainty Calculation}

{First of all, the filtered value ${\Delta _f}\left( t \right)$ of uncertainty $\Delta \left( t \right)$ is to be obtained from the error equation \eqref{eq10}.} For this purpose, we introduce the following aperiodic links with resetting:
\begin{equation}\label{eq11}
{\small
\begin{array}{c}
{{\dot \mu }_f}\left( t \right) =  - k{\mu _f}\left( t \right) + {{\dot e}_{ref}}\left( t \right){\rm{,}}\;{{\mu _f}\left( {t_r^ + } \right) = {0_n}{\rm{,}}}\\
{{\dot e}_f}\left( t \right) =  - k{e_f}\left( t \right) + {e_{ref}}\left( t \right){\rm{,}}\;{{e_f}\left( {t_r^ + } \right) = {0_n}{\rm{,}}}
\end{array}
}
\end{equation}
\begin{equation}\label{eq12}
{\small
\begin{array}{c}
{{\dot u}_{adf}}\left( t \right) =  - k{u_{adf}}\left( t \right) + {u_{ad}}\left( t \right){\rm{,}}\;{{u_{adf}}\left( {t_r^ + } \right) = {0_m}{\rm{,}}}\;\\
{{\dot \Phi }_f}\left( {x\left( t \right)} \right) =  - k{\Phi _f}\left( {x\left( t \right)} \right) + \Phi \left( {x\left( t \right)} \right){\rm{,}}\;{{\Phi _f}\left( {x\left( {t_r^ + } \right)} \right) = {0_p}{\rm{,}}}
\end{array}
}
\end{equation}
{where $t_r^ +  = {t_k}$ is the resetting time instant.}

\begin{lemma}\label{lemma1}{
Let $k > 0$ be sufficiently large, then the filtered value ${\Delta _f}\left( t \right) \in {\mathbb{R}^m}$ of uncertainty $\Delta \left( t \right)$ can be evaluated as follows:
\begin{equation}\label{eq13}
{\small
\begin{array}{c}
{\Delta _f}\left( t \right) = {B^{\rm{\dag }}}\left( {{e_{ref}}\left( t \right) - k{e_f}\left( t \right) - {e^{ - k\left( {t - t_r^ + } \right)}}{e_{ref}}\left( {t_r^ + } \right)}- \right.\\
 - \left. {{A_{ref}}{e_f}\left( t \right) + B{u_{adf}}\left( t \right)} \right) = {\Theta ^{\rm{T}}}\left( t \right){\Phi _f}\left( {x\left( t \right)} \right).
\end{array}
}
\end{equation}
}
{Proof can be found in the Supplementary Material \cite{b24}.}
\end{lemma}

\begin{proposition}\label{proposition1}
A sufficient condition of ${\Phi _f}\left( x \right) \in {\rm{FE}}$ is that $k > 0$ and $\Phi \left( x \right) \in {\rm{FE}}$ (for proof see Lemma 6.8 in \cite{b6}).
\end{proposition}

\subsection{Scalarization via DREM}

{The DREM procedure \cite{b25,b26,b27,b28} relaxes PE condition and transforms ${\Phi _f}\left( x \right)$ into a new scalar regressor $\omega \left( t \right) \in \mathbb{R}.$ This allows one to both simplify significantly the synthesis of the adaptive law with the required properties and improve the transient quality of the obtained estimates (see \cite{b25,b26,b27,b28} for more details). DREM consists of extension and mixing steps. The extended linear regression equation is derived as a result of the first of them:}
\begin{equation}\label{eq14}
\begin{array}{c}
\dot y\left( t \right) =  - ly\left( t \right) + {\Phi _f}\left( {x\left( t \right)} \right)\Delta _f^{\rm{T}}\left( t \right){\rm{,\;}}{y\left( {t_r^ + } \right) = {0_{p \times m}}{\rm{,}}}\\
\dot \varphi \left( t \right) =  - l\varphi \left( t \right) + {\Phi _f}\left( {x\left( t \right)} \right)\Phi _f^{\rm{T}}\left( {x\left( t \right)} \right){\rm{,}}\;{\varphi \left( {t_r^ + } \right) = {0_{p \times p}}}.
\end{array}
\end{equation}

The mixing step is represented as the following lemma.
\begin{lemma}\label{lemma2}{
Let $l > 0$ be sufficiently large, then from \eqref{eq14} the mixed regression can be obtained as follows:
\begin{equation}\label{eq15}
{\small
Y\!\left( t \right)\! =\! adj\left\{ {\varphi \left( t \right)} \right\}y\!\left( t \right)\! =\! det \left\{ {\varphi \left( t \right)} \right\}{I_{p \times p}}\Theta\! \left( t \right)\! =\! \omega \!\left( t \right)\Theta \!\left( t \right),}
\end{equation}
where $Y\left( t \right) \in {\mathbb{R}^{p \times m}}.$}

Proof can be found in the Supplementary Material \cite{b24}.
\end{lemma}

\begin{proposition}\label{proposition2}
If ${\Phi _f}\left( x \right)$ is {\rm FE}, then $\omega \left( t \right)$ is also {\rm FE} [27].
\end{proposition}

The equations of the conventional estimator of the regression \eqref{eq15} parameters are written as: $\dot{ \tilde \Theta} \left( t \right) =  - \gamma {\omega ^2}\left( t \right)\tilde \Theta \left( t \right){\rm{,}}$ where $\gamma  > 0$. According to the results of \cite{b25,b26,b27,b28}, the following implication holds for this adaptive law: $\omega \left( t \right) \notin {L_2} \Leftrightarrow \mathop {\lim }\limits_{t \to \infty } \tilde \Theta \left( t \right) = 0.$ Since, when $\omega \left( t \right) \notin {L_2}$, only the asymptotic convergence of the parameter error is guaranteed, the application of the DREM procedure alone is insufficient to achieve the stated goal.

\subsection{Resetting Filtration and Adaptive Law Derivation}

Then the integral filter with exponential forgetting and resetting should be used \cite{b29} to guarantee the exponential convergence of the error $\tilde \Theta \left( t \right)$ when $\omega \left( t \right) \in {\rm{FE}}$:
\begin{equation}\label{eq16}
{\dot v_f}\left( t \right) = {\rm{exp}}\left( { - \int\limits_{t_r^ + }^t {\sigma d\tau } } \right)v\left( t \right){\rm{,}}\;{v_f}\left( {t_r^ + } \right) = 0,
\end{equation}
where $\sigma  > 0$, $v\left( t \right)$ and ${v_f}\left( t \right)$ are the input and output respectively.

\begin{remark}\label{remark2}
According to \cite{b20}, the filtration \eqref{eq16} allows one to obtain a non-zero regressor and thereby, as proved in \cite{b20}, relax the PE requirement for exponential convergence of the parameter error. In contrast to other methods, e.g. the integral-like filter \cite{b18}, the filter \eqref{eq16} guarantees a bounded ${v_f}\left( t \right)$ in case of noise and disturbances.
\end{remark}

If the input of ${v_f}\left( t \right)$ is $\omega \left( t \right)Y\left( t \right)$, then, using \eqref{eq15}, we obtain:
\begin{equation}\label{eq17}
{\small
\begin{array}{c}
\Upsilon \!\left( t \right)\! =\! \int\limits_{t_r^ + }^t {{\rm{exp}}\!\left( { - \int\limits_{t_r^ + }^t {\sigma d{\tau _1}} } \right)\omega\! \left( \tau  \right)Y\!\left( \tau  \right)d\tau } {\rm{,}}\;\Upsilon\! \left( {t_r^ + } \right)\! =\! {0_{p \times m}}{\rm{,}}\\
\Omega \left( t \right) = \int\limits_{t_r^ + }^t {{\rm{exp}}\left( { - \int\limits_{t_r^ + }^t {\sigma d{\tau _1}} } \right){\omega ^2}\left( \tau  \right)d\tau } {\rm{,}}\;\Omega \left( {t_r^ + } \right) = 0,
\end{array}
}
\end{equation}
where $\Upsilon \left( t \right) \in {R^{p \times m}}$.

According to Assumption \ref{assumption3}, if $t = {t_k}$, then $\Phi \left( x \right) \in {\rm{FE}}$ for $\left[ {{t_k}{\rm{;}}\;{t_k} + T} \right]$. According to Propositions \ref{proposition1} and \ref{proposition2}, if $\Phi \left( x \right) \in {\rm{FE}}$, then $\omega \left( t \right) \in {\rm{FE}}$. So, taking into account the definition of ${t_k}$ and $t_r^ + $, $\omega \left( t \right) \in {\rm{FE}}$ over the interval $\left[ {t_r^ + {\rm{;}}\;{t_e}} \right]{\rm{,}}\;{t_e} \ge t_r^ + $. So, the following proposition can be introduced and proved.

\begin{proposition}\label{proposition3}
If $\omega \left( t \right) \in {\rm{FE}}$ over the interval $\left[ {t_r^ + {\rm{;\;}}{t_e}} \right]$, then
\begin{enumerate}
    \item[1)] $\forall t \ge t_r^ + {\rm{\;}}\Omega \left( t \right) \in {L_\infty },{\rm{\;}}\Omega \left( t \right) \ge 0$;
    \item[2)] $\forall t \ge {t_e}{\rm{\;}}\Omega \left( t \right) > 0,{\rm{\;}}{\Omega _{LB}} \le \Omega \left( t \right) \le {\Omega _{UB}}$.
\end{enumerate}

For proof, please, refer to the Supplementary Material \cite{b24}.
\end{proposition}

According to Proposition \ref{proposition3} and in contrast to $\omega \left( t \right)$, the regressor $\Omega \left( t \right)$, which is obtained by filtering with resetting \eqref{eq16}, does not vanish over any time interval between two consecutive changes of the reference value. It allows us to derive the law to adjust the adaptive controller parameters ${u_{ad}}\left( t \right)$ according to the composite method \cite{b5}:
\begin{equation}\label{eq18}
{\footnotesize
\begin{array}{c}
\dot {\hat \Theta}\! \left( t \right)\! =\! {\Gamma _1}\!\Phi\! \left( {x\left( t \right)} \right)\!e_{ref}^{\rm T}\!\left( t \right)PB\! +\! {\Gamma _2}\!\left( t \right)\Omega\! \left( t \right)\!\left( {\Upsilon\! \left( t \right)\! -\! \Omega\! \left( t \right)\hat \Theta\! \left( t \right)} \right),\\
{{\dot \Gamma }_2}\left( t \right) = {\lambda _1}{\Gamma _2}\left( t \right) - {\lambda _2}\Gamma _2^2\left( t \right){\Omega ^2}\left( t \right),{\rm{}}\;{\Gamma _2}\left( {t_{0}} \right) > 0,
\end{array}
}
\end{equation}
where ${\Gamma _1} \in {\mathbb{R}^{p \times p}}$ and ${\Gamma _2}\left( t \right) \in \mathbb{R}$ are the adaptive gains, ${\lambda _1} > 0$, ${\lambda _2} > 0$ are the forgetting and damping factors respectively.

In \eqref{eq18} the first summand is to provide the convergence of the tracking error ${e_{ref}}\left( t \right)$, and the second one – the parameter error $\tilde \Theta \left( t \right)$. The properties of $\Omega \left( t \right)$, which are proved in Proposition \ref{proposition3}, give the opportunity to use variable adaptive gain ${\Gamma _2}\left( t \right)$ in \eqref{eq18}.

\begin{proposition}\label{proposition4}
If $\omega \left( t \right) \in {\rm{FE}}$ over the interval $\left[ {t_r^ + {\rm{;}}\;{t_e}} \right]$, then \linebreak  $\forall t > t_r^ + $ the following inequality holds: ${\Gamma _{2\min }} \le {\Gamma _2}\left( t \right) \le {\Gamma _{2\max }}$.

For proof, please, refer to the Supplementary Material \cite{b24}.
\end{proposition}

\begin{remark}\label{remark3}
According to \cite{b20,b28,b29}, for the boundedness of ${\Gamma _2}\left( t \right)$, it is sufficient that $\Omega \left( t \right)$ is bounded by its lower ${\Omega _{LB}}$ and upper ${\Omega _{UB}}$ bounds, both of which are above zero, and does not converge to zero $\Omega \bcancel{ \to }0$. Therefore, the proof of Proposition \ref{proposition4} \cite{b24} is given in brief, but sufficient version. We recommend to refer to \cite{b20,b28,b29} for more details.
\end{remark}

\begin{remark}\label{remark4}
Right-hand side of \eqref{eq18} has a point of discontinuity due to filtering \eqref{eq16}. However, the time instant $t_r^ + $, at which the discontinuity occurs, is determined by the external signal $r\left( t \right)$ and does not depend on the internal signals $x\left( t \right)$ and ${e_{ref}}\left( t \right)$ of the closed-loop system \eqref{eq10}. This completely excludes the possibility of chattering in \eqref{eq16}, \eqref{eq18} and allows analyzing the stability of the closed-loop \eqref{eq10}, assuming the time instant $t_r^ + $ is initial one.
\end{remark}

\section{Stability Analysis}

The scope of this section is to analyze the stability of the system \eqref{eq10} when the law \eqref{eq18} is applied to adjust ${u_{ad}}\left( t \right)$. In this case, the behavior of the error $\xi \left( t \right)$ is completely defined by the mutual relation of ${t_j}$ and the interval $\left[ {t_r^ + {\rm{;}}\;{t_e}} \right]$. Three situations are possible: 1) ${t_j} \le t_r^ + $ – the unknown parameters change has happened before the change of the reference value $r\left( t \right)$; 2) ${t_j} \in \left( {t_r^ + {\rm{;}}\;{t_e}} \right)$ – the unknown parameters change has happened when the regressor is FE because of $r\left( t \right)$ value change; 3) ${t_j} \ge {t_e}$ – the unknown parameters change has happened after FE time range. The boundedness of ${\Gamma _2}\left( t \right)$ for all cases allows us to introduce the Lyapunov function:
\begin{equation}\label{eq19}
{\footnotesize
\begin{array}{c}
V\!\left( \xi  \right)\! =\! e_{ref}^{\rm T}P{e_{ref}}\! +\! {\rm{tr}}\!\left( {{{\tilde \Theta }^{\rm T}}\!\Gamma _1^{ - 1}\tilde \Theta } \right){\rm{\!,\!}}\;H \!=\! blockdiag\left\{ {P\!,\!{\rm{}}\;\Gamma _1^{ - 1}} \right\}{\rm{\!,}}\\
{\rm{}}\;\underbrace {{\lambda _{\min }}\left( H \right)}_{{\lambda _{\mathop{\rm m}\nolimits} }}{\left\| \xi  \right\|^2} \le V\left( {\left\| \xi  \right\|} \right) \le \underbrace {{\lambda _{\max }}\left( H \right)}_{{\lambda _M}}{\left\| \xi  \right\|^2}.
\end{array}
}
\end{equation}

The derivative of \eqref{eq19} with respect to the equations of the system \eqref{eq10} and the adaptive loop \eqref{eq18} for any $t \ne {t_j}$ is written as:
\begin{equation}\label{eq20}
\dot V\left( \xi  \right) =  - e_{ref}^{\rm T}Q{e_{ref}} - 2{\rm{tr}}\left( {{{\tilde \Theta }^{\rm T}}\Gamma _1^{ - 1}{\Gamma _2}\Omega \left( {\Upsilon  - \Omega \hat \Theta } \right)} \right),
\end{equation}

For further analysis, let each case be considered separately.
\begin{theorem}\label{theorem1}
(Case 1) If ${t_j} \le t_r^ + $ and $\Phi \left( x \right) \in {\rm{FE}}$, then the augmented error $\xi \left( t \right)$ is exponentially stable $\left( {\xi \left( t \right) \in {\rm{GES}}} \right)$.
\end{theorem}

\begin{theorem}\label{theorem2}
(Case 2) If ${t_j} \in \left( {t_r^ + {\rm{;}}\;{t_e}} \right)$ and $\Phi \left( x \right) \in {\rm{FE}}$, then the augmented error $\xi \left( t \right)$ is exponentially ultimately bounded $\left( {\xi \left( t \right) \in {\rm{EUB}}} \right)$ with the ultimate bound $R$.
\end{theorem}

\begin{theorem}\label{theorem3}
(Case 3) If ${t_j} \ge {t_e}$ and $\Phi \left( x \right) \in {\rm{FE}}$, then the augmented error $\xi \left( t \right)$ is exponentially ultimately bounded $\left( {\xi \left( t \right) \in {\rm{EUB}}} \right)$ with ultimate bound ${R_1}$.

Proofs of Theorems 1-3 are given in Supplementary Material \cite{b24}.
\end{theorem}

It follows from the proofs that if the unknown parameters $\Theta \left( t \right)$ change their values: 1) {\emph{before}} the finite excitation, caused by the change of the reference $r\left( t \right)$, then $\xi \left( t \right) \in {\rm{GES}}$; 2) {\emph{during}} or {\emph{after}} FE time range, then $\xi \left( t \right) \in {\rm{EUB}}$ with ultimate bound $R$ or ${R_1}$ \cite{b24}.

For cases 2 and 3 let the ways to minimize the ultimate bound of the errors ${e_{ref}}\left( t \right)$ and $\tilde \Theta \left( t \right)$ be analyzed separately for each of them. For this purpose, taking into account the definitions of ${\kappa _{min}}{\rm{,}}\;\varepsilon {\rm{,}}\;{\varepsilon _1}$ \cite{b24} and assuming \linebreak ${\lambda _{\min }}\left( {{\Gamma _1}} \right) = {\lambda _{\max }}\left( {{\Gamma _1}} \right)$, the equations of bounds of both errors ${e_{ref}}\left( t \right)$ and $\tilde \Theta \left( t \right)$ are obtained from proofs of Theorems \ref{theorem2} and \ref{theorem3}:
\begin{equation}\label{eq21}
{\scriptsize
\begin{array}{l}
\left\| {\tilde \Theta } \right\| \!\le\! \left\{ \!\!\!\!\begin{array}{l}
{\overline r_1}\!\! =\! \!\sqrt {{{\!\left( {\frac{{{\Omega _{1UB}}{\Omega _{UB}}{\Gamma _{2\max }}}}{{\Omega _{LB}^2{\Gamma _{2\min }}}}} \right)}^2}\!\!\!\left\| {{\Theta _1}} \right\|_F^2} {\rm{,\!\;for\;}}{t_j} \!\!\in\!\! \left( {t_r^ + {\rm{; }}{t_e}} \right)\!\!{\rm{,}}\\
{\overline r_2} = \sqrt {{{\left( {\frac{{{\Gamma _{2\max }}\Omega _{UB}^2}}{{{\Gamma _{2\min }}\Omega _{LB}^2}}} \right)}^2}\!\!\left\| {{\Theta _1}} \right\|_F^2} {\rm{,\;for\;}}{t_j} \ge {t_e}{\rm{,}}
\end{array} \right.\\
\left\| {{e_{ref}}} \right\| \le\\
\!\!\! \left\{\!\!\!\! \begin{array}{l}
{\overline r_1}\!\sqrt{\!{\lambda _{\min }}\!\!\left(\! {\Gamma _1^{ - 1}} \!\right)\!{\!\Gamma _{2\!\min }}\Omega _{LB}^2\frac{{{\lambda _{\max }}\!\left( P \right)}}{{{\lambda _{\min }}\!\left( P \right)}}\!\lambda _{\min }^{ - 1}\!\!\left( Q \right)} {\rm{,\!\;for\;}}{t_j}\!\! \in\!\! \left( {t_r^ + {\rm{;}}{t_e}}\! \right)\!\!{\rm{,}}\\
{\overline r_2}\sqrt {{\lambda _{\min }}\left(\! {\Gamma _1^{ - 1}}\! \right)\!{\Gamma _{2\!\min }}\Omega _{LB}^2\frac{{{\lambda _{\max }}\left( P \right)}}{{{\lambda _{\min }}\left( P \right)}}\lambda _{\min }^{ - 1}\left( Q \right)} {\rm{,\!\;for\;}}{t_j}\!\! \ge\!\! {t_e}{\rm{,}}
\end{array} \right.
\end{array}
}
\end{equation}

It follows from \eqref{eq21} that in both cases the ultimate bound of the parameter error can be reduced by the minimization of the multiplication ${\Gamma _{2\max }}\Gamma _{2\min }^{ - 1}$. However, such minimization is often difficult to be implemented in practice, because the values of ${\Gamma _{2\max }}$ and ${\Gamma _{2\min }}$, according to Proposition \ref{proposition4}, are determined by the regressor $\Omega \left( t \right)$. The ultimate bound of the tracking error ${e_{ref}}\left( t \right)$ can be reduced by improvement of ${\lambda _{\max }}\left( {{\Gamma _1}} \right)$ or reduction of ${\Gamma _{2\min }}$. But the improvement of ${\lambda _{\max }}\left( {{\Gamma _1}} \right)$ leads to the higher sensitivity of the adaptive law \eqref{eq18} to noise and disturbances and deterioration of the quality of the transient process of $\dot {\hat \Theta} \left( t \right)$ and ${u_{ad}}\left( t \right)$. The reduction of ${\Gamma _{2\min }}$ (by the change of ${\lambda _1}$ and ${\lambda _2}$) results in the decrease of the convergence speed to the ideal parameters of the uncertainty in case 1. So, in practice, ${\lambda _{\max }}\left( {{\Gamma _1}} \right)$, ${\lambda _1}$ and ${\lambda _2}$ must be chosen by compromise between the ultimate bounds $R{\rm{,}}\;{R_1}$ and the quality of the transient process of $\dot {\hat \Theta} \left( t \right)$ and ${u_{ad}}\left( t \right)$.

Thus, according to the conducted stability analysis, the obtained adaptive law \eqref{eq18} provides the properties required by Goal when the uncertainty parameters change before the reference $r\left( t \right)$ switch.

\section{Comparison with Known Adaptive Laws}

The main difference of the developed system from the majority of the known ones is the requirement \eqref{eq6} of a certain type of the reference signal $r\left( t \right)$, which is necessary to guarantee the stable implementation of the resetting procedure for the filtering \eqref{eq11}, \eqref{eq12}, \eqref{eq14}, \eqref{eq17}. This fact does not allow one to apply the obtained system in cases when $r\left( t \right)$ is the output of the command filter or the outer loop controller. But, as far as the plants, for which $r\left( t \right)$ satisfies the requirement stated in Assumption \ref{assumption2}, are concerned, the obtained system allows one to provide the exponential convergence of the estimates of the piecewise-constant uncertainty parameters to their ideal values when such values have changed before the change of the reference $r\left( t \right)$ and the regressor is FE.

In contrast to existing CMRAC schemes, in this paper, owing to the application of the DREM procedure, it is proposed \eqref{eq18} to augment the basic adjustment law $\Phi \left( x \right)e_{ref}^{\rm{T}}PB$ with not a matrix law, which ensures convergence of the integral error of the uncertainty identification, but with a scalar DREM-based one to provide monotonic estimation of each $\tilde \Theta \left( t \right)$ element. {So, being used together with $\Phi \left( x \right)e_{ref}^{\rm{T}}PB$, it does not cause additional fluctuations of the transient curves of $\hat \Theta \left( t \right)$ values. Moreover, if the following condition holds: ${\Gamma _1}\Phi \left( {x\left( t \right)} \right)e_{ref}^{\rm{T}}\left( t \right)PB =$ \linebreak $= o\left( {{\Gamma _2}\left( t \right)\Omega \left( t \right)\left( {\Upsilon \left( t \right) - \Omega \left( t \right)\hat \Theta \left( t \right)} \right)} \right)$, then the proposed adaptive law \eqref{eq18} ensures that $\left| {{{\tilde \Theta }_i}\left( {{t_a}} \right)} \right| \le \left| {{{\tilde \Theta }_i}\left( {{t_b}} \right)} \right|{\rm{,\;}}\forall {t_a} \ge {t_b}$.}

The properties of the regressor $\Omega \left( t \right)$, which is obtained using the filtration with resetting \eqref{eq16}, also make it possible, unlike in other CMRAC schemes, to use a variable gain ${\Gamma _2}\left( t \right)$ in \eqref{eq18}.

Next, we briefly compare the developed law \eqref{eq18} with some previously proposed in the literature.

\subsection{Comparison with Basic Robust Adaptive law}
In practice, the conventional adaptive law $\Phi \left( x \right)e_{ref}^{\rm{T}}PB$ is always augmented with the robust modifications. For example, if the $\sigma$-modification is used, then it is written as \cite{b6, b9}:
\begin{equation}\label{eq22}
\dot {\hat \Theta} \left( t \right) = {\Gamma _1}\left( {\Phi \left( {x\left( t \right)} \right)e_{ref}^{\rm T}\left( t \right)PB - \sigma \hat \Theta \left( t \right)} \right)
\end{equation}

The robust adaptive law \eqref{eq22} guarantees $\xi \left( t \right) \in {\rm{EUB}}$ for the plant \eqref{eq2} in the presence of the bounded disturbances. As for the proposed adaptive law \eqref{eq18}, it is proved in the theorems \ref{theorem1}-\ref{theorem3} that $\xi \left( t \right) \in {\rm{GES}}$ or $\xi \left( t \right) \in {\rm{EUB}}$ in the absence of the disturbances. This is both necessary and sufficient to provide robustness $\left( {\xi \left( t \right) \in {\rm{EUB}}} \right)$ of \eqref{eq18} against bounded disturbances \cite{b6,b8,b9}. So, in contrast to the conventional adaptive law $\Phi \left( x \right)e_{ref}^{\rm{T}}PB$, not only does the proposed one \eqref{eq18} guarantee $\xi \left( t \right) \in {\rm{GES}}$ in case 1, but also it does not require additional robust modifications.

\subsection{Comparison with Switched MRAC}

Using \eqref{eq12}, \eqref{eq13}, and \eqref{eq14}\footnote{However, in \cite{b22, b30} the above-mentioned filtration is used without resetting.}, it is proposed in \cite{b30} to apply the following switched adaptive law:
\begin{equation}\label{eq23}
{\small
\begin{array}{c}
\dot {\hat \Theta} \left( t \right) = {\Gamma _1}\left( {{T_e} + {T_l} + {T_{ll}} + {T_{sw}}} \right)\\
{T_e} = \Phi \left( {x\left( t \right)} \right)e_{ref}^{\rm T}\left( t \right)PB{\rm{;}}\;{{\mathop{\rm T}\nolimits} _l} = {k_l}{\left( {{\Delta _f}\left( t \right)\!-\!{{\hat \Theta }^{\rm T}}{\Phi _f}\left( {x\left( t \right)} \right)} \right)^{\rm T}}\!\!{\rm{;}}\\
{{\mathop{\rm T}\nolimits} _{ll}} = {k_{ll}}\left( {y\left( t \right) - \varphi \left( t \right)\hat \Theta \left( t \right)} \right){\rm{;}}\;\\
{T_{sw}} = {k_{sw}}\left( {{y_{sw}}\left( t \right) - {\varphi _{sw}}\left( t \right)\hat \Theta \left( t \right)} \right).
\end{array}
}
\end{equation}

Here ${k_l} > 0,{\rm{}}\;{k_{ll}} > 0,{\rm{}}\;{k_{sw}} > 0,{\rm{\;}}{y_{sw}}\left( t \right) \in {\mathbb{R}^{p \times m}}$ and \linebreak ${\varphi _{sw}}\left( t \right) \in {\mathbb{R}^{p \times p}}$ are defined as:
\begin{equation}\label{eq24}
\begin{array}{l}
{y_{sw}}\left( t \right) = \left\{ \begin{array}{l}
0_{p \times m},{\rm{ if \;}} det\left( {\int\limits_0^t {{\Phi _f}\left( \tau  \right)\Phi _f^{\rm T}\left( \tau  \right)d\tau } } \right) = 0,\\
y\left( T \right){\rm{\;otherwise}}{\rm{,}}
\end{array} \right.\\
{\varphi _{sw}}\left( t \right) = \left\{ \begin{array}{l}
0_{p \times p},{\rm{ if \;}} det\left( {\int\limits_0^t {{\Phi _f}\left( \tau  \right)\Phi _f^{\rm T}\left( \tau  \right)d\tau } } \right) = 0,\\
\varphi \left( T \right){\rm{\;otherwise}}{\rm{.}}
\end{array} \right.
\end{array}
\end{equation}
where $T$ is a time instant, when the determinant in \eqref{eq24} becomes positive, $\varphi \left( T \right)$ is a filtered regressor of full rank.

According to the proof in \cite{b30}, the law \eqref{eq24} guarantees \linebreak $\xi \left( t \right) \in {\rm{GES}}$ when $\Phi \left( x \right) \in {\rm{IE}}$. As ${\Phi _f}\left( x \right)\Phi _f^{\rm{T}}\left( x \right)$ is the positive semi-definite matrix, then the switching in \eqref{eq24} is possible only once. Consequently, in contrast to \eqref{eq18}, the adaptive law \eqref{eq24} guarantees $\xi \left( t \right) \in {\rm{GES}}$ only when the uncertainty parameters are constant.

\subsection{Comparison with FE CMRAC}

The FE CMRAC adaptive law \cite{b22} also uses filtering, which is similar to \eqref{eq11}, \eqref{eq12}$^{1}$, to obtain the numerical value of the uncertainty \eqref{eq13}. It also uses the variable forgetting factor $l\left( t \right)$ of the Kreisselmeier filter \eqref{eq14}:
\begin{equation}\label{eq25}
l\left( t \right) = {l_m} + \left( {{l_M} - {l_m}} \right)\tanh \left( {\vartheta \left\| {{{\dot \Phi }_f}\left( x \right)} \right\|} \right),
\end{equation}
where ${l_m}$ and ${l_M}$ are minimum and maximum values of the parameter $l\left( t \right)$, $\vartheta  > 0$, $tanh\left(.\right)$ is the hyperbolic tangent function. The law \eqref{eq25} to adjust $l\left( t \right)$ allows one to set the higher weight to the uncertainty data, which are obtained when ${\Phi _f}\left( x \right)$ changes rapidly.

The main point of the novelty of the study \cite{b22} is the developed algorithm to obtain the full-rank regressor:
\begin{equation}\label{eq26}
\begin{array}{c}
{t_a} = {\rm{max}}\left\{ {\mathop {{\rm{arg max \;}}}\limits_{\tau  \in \left[ {0{\rm{;\;t}}} \right]} {\lambda _{\min }}\left( {\varphi \left( \tau  \right)} \right)} \right\},\\
{\varphi _a}\left( t \right) = \varphi \left( {{t_a}} \right),{\rm{}}\;{y_a}\left( t \right) = y\left( {{t_a}} \right).
\end{array}
\end{equation}

It allows one to choose the adaptive law as:
\begin{equation}\label{eq27}
{\footnotesize
\dot {\hat \Theta}  = {\Gamma _1}\left( {\Phi \left( {x\left( t \right)} \right)e_{ref}^{\rm T}\left( t \right)PB\! +\! {\Gamma _2}\left( {{y_a}\left( t \right)\! -\! {\varphi _a}\left( t \right)\hat \Theta \left( t \right)} \right)} \right)
}
\end{equation}

In contrast to switching \eqref{eq24} in Switched MRAC, algorithm \eqref{eq26} allows obtaining a full-rank regressor every time when FE exists. However, the values of ${\varphi _a}\left( t \right)$ and ${y_a}\left( t \right)$ will be updated according to \eqref{eq26} only if the minimum eigenvalue of the matrix $\varphi \left( t \right)$ over the new FE time range is higher than it has been over the previous one.

In addition, in contrast to the filtering with the resetting procedure \eqref{eq17} used in this study, the Kreisselmeier filtration \eqref{eq14} with a variable parameter \eqref{eq25} does not allow one to forget completely the outdated information about $\varphi \left( t \right)$ and $y\left( t \right)$. Therefore, the adaptive law \eqref{eq27} is applicable to piecewise-constant uncertainty parameters only if: 1) a new finite excitation leads to an improvement of the minimum eigenvalue of $\varphi \left( t \right)$, and 2) $y\left( t \right)$ has been completely forgotten by the moment the new finite excitation occurs. In this case, the proposed adaptation loop \eqref{eq18}, according to the proof of Theorem \ref{theorem1}, guarantees $\xi \left( t \right) \in {\rm{GES}}$ without additional conditions of complete forgetting of $y\left( t \right)$ and growth of the minimal eigenvalue of $\varphi \left( t \right)$.

\subsection {Comparison with Directional Forgetting CL MRAC}

In \cite{b21} it was proposed to use directional forgetting in \eqref{eq14}\footnote{In \cite{b21, b22i5} the above-mentioned filtration is used without resetting.}. This allowed one to implement forgetting of the outdated data $y\left( t \right)$ only in the direction of the newly obtained data. Further, the authors used a switching algorithm, based on the rank condition, between filtering with the directional forgetting and the one with the open-loop integrator \cite{b18}:
\begin{equation}\label{eq28}
{\footnotesize
\begin{array}{l}
\dot y\left( t \right) = \\
\left\{\!\!\!\! \begin{array}{l}
{\Phi _f}\left( t \right)\Delta _f^{\rm T}\left( t \right)\!{\rm{,\;\!if\;\! rank}}\left( {\varphi \left( t \right)} \right)\!\! <\!\! {\rm{rank}}\!\left(\! {\varphi \left( t \right) + {\Phi _f}\left( t \right)\Phi _f^{\rm T}\left( t \right)}\! \right)\!\!{\rm{,}}\\
 - l\frac{{\varphi \left( t \right){\Phi _f}\left( t \right)\Phi _f^{\rm T}\left( t \right)}}{{\Phi _f^{\rm T}\left( t \right)\varphi \left( t \right){\Phi _f}\left( t \right)}}y\left( t \right)\! +\! {\Phi _f}\left( t \right)\Delta _f^{\rm T}\left( t \right){\rm{ otherwise}}{\rm{,}}
\end{array} \right.\\
\dot \varphi \left( t \right) = \\
\left\{\!\!\!\! \begin{array}{l}
{\Phi _f}\left( t \right)\Phi _f^{\rm T}\left( t \right)\!{\rm{,\;\!if\;\!rank}}\left( {\varphi \left( t \right)} \right)\!\! < \!\!{\rm{rank}}\!\left( {\varphi \left( t \right)\! +\! {\Phi _f}\left( t \right)\Phi _f^{\rm T}\left( t \right)} \right)\!\!{\rm{,}}\\
 - l\frac{{\varphi \left( t \right){\Phi _f}\left( t \right)\Phi _f^{\rm T}\left( t \right)}}{{\Phi _f^{\rm T}\left( t \right)\varphi \left( t \right){\Phi _f}\left( t \right)}}\varphi \left( t \right) + {\Phi _f}\left( t \right)\Phi _f^{\rm T}\left( t \right){\rm{ otherwise}}{\rm{.}}
\end{array} \right.
\end{array}
}
\end{equation}

It allows choosing the adaptive law as:
\begin{equation}\label{eq29}
{\footnotesize
\dot {\hat \Theta}\! \left( t \right)\! =\! {\Gamma _1}\!\left(\! {\Phi \left( {x\left( t \right)} \right)e_{ref}^{\rm{T}}\!\left( t \right)\!PB\! +\! {\Gamma _2}\!\left( {y\left( t \right)\! -\! \varphi \left( t \right)\hat \Theta \left( t \right)}\! \right)} \right).    
}
\end{equation}

The common disadvantage of the proposed method and the directional forgetting CL MRAC one is the need to meet the special conditions, under which the exponential convergence of the parameter error to zero is guaranteed. The resetting filtration method requires that the uncertainty parameters must change their values before the change of the reference $r\left( t \right)$. The directional forgetting CL MRAC approach requires stricter requirement ${\theta _j}\tilde \Theta^{\rm{T}} \left( t \right) \ge 0$ to be met. However, the function of $r\left( t \right)$ change is defined by a user of the adaptive system, so the condition that the parameters change their values in prior to the reference $r\left( t \right)$ is weaker then ${\theta _j}\tilde \Theta^{\rm{T}} \left( t \right) \ge 0$. 

\subsection{{Comparison with efficient learning MRAC}} {
In \cite{b22i5} an algorithm to calculate the filtered uncertainty value on the basis of aperiodic filtering, which is similar to \eqref{eq11} and \eqref{eq12}$^{2}$, is also used. However, unlike \eqref{eq25} and \eqref{eq28}, the filter \eqref{eq14}$^{2}$ forgetting factor is proposed to be defined as follows:
\begin{equation}\label{eq29i5}
l\left( t \right) = \left\{\!\!\! \begin{array}{l}
{l_0}{\rm{,\;if\;}}{\textstyle{{2{\lambda _{\min }}\left( {\varphi \left( t \right)} \right) - \lambda _{\min }^{UB} - \lambda _{\min }^{LB}} \over {\lambda _{\min }^{UB} - \lambda _{\min }^{LB}}}} \ge 1\\
{\textstyle{{{l_0}} \over 2}}\left( {{\textstyle{{2{\lambda _{\min }}\left( {\varphi \left( t \right)} \right) - \lambda _{\min }^{UB} - \lambda _{\min }^{LB}} \over {\lambda _{\min }^{UB} - \lambda _{\min }^{LB}}}} + 1} \right){\rm{,\;otherwise}}
\end{array} \right.
\end{equation}
where $0 < \lambda _{\min }^{LB} \le {\lambda _{\min }}\left( {\varphi \left( t \right)} \right) \le \lambda _{\min }^{UB}$ are the minimum eigenvalue of the regressor and its lower and upper bounds, ${l_0} > 0$ is the scaling factor.}
{
Based on the filtration \eqref{eq14} with variable forgetting factor \eqref{eq29i5}, in \cite{b22i5} it is proposed to use the composite law in the form of \eqref{eq29}. According to \eqref{eq29i5}, if the eigenvalue ${\lambda _{\min }}\left( {\varphi \left( t \right)} \right) \to \lambda _{\min }^{UB}$, then $l\left( t \right) \to {l_0}$, and filtering \eqref{eq14} with \eqref{eq29i5} provides a high update rate of data on the uncertainty. On the contrary, if the eigenvalue ${\lambda _{\min }}\left( {\varphi \left( t \right)} \right) \to \lambda _{\min }^{LB}$, then $l\left( t \right) \to 0$, and filtering \eqref{eq14} loses the ability to update the data on the uncertainty in the limit. Thus, when  \eqref{eq29i5} is used in \eqref{eq14}, the law \eqref{eq29} guarantees the convergence of the parameter error if after the time instant ${t_j}$ the minimum eigenvalue ${\lambda _{\min }}\left( {\varphi \left( t \right)} \right)$ keeps value, which is close to $\lambda _{\min }^{UB}$, over a sufficiently long time range. Compared to the requirement that the uncertainty parameters switch is prior to the reference signal change, such a condition seems to be restrictive, difficult to be satisfied for many important cases, and is usually equivalent to the PE requirement.}
{\begin{remark}\label{remark5}
In order to illustrate the advantages of proposed approach systematically it should be noted that its main salient feature is that the outdated data are removed with the help of resetting procedure driven by exogenous reference signal without dependence on any internal signals of the closed loop, therefore, the developed method completely excludes the possibility of chattering and avoid superpositional mixing instantaneously (not aperiodically).
\end{remark}}

\section{Numerical Simulation}
The wing-rock phenomenon in the roll motion of slender delta wings has been chosen as the plant for the experiments:
\begin{equation}\label{eq30}
{\begin{bmatrix}
{{{\dot x}_1}\left( t \right)}\\
{{{\dot x}_2}\left( t \right)}
\end{bmatrix}}\! =\! {\begin{bmatrix}
0&1\\
0&0
\end{bmatrix}}{\begin{bmatrix}
{{x_1}\left( t \right)}\\
{{x_2}\left( t \right)}
\end{bmatrix}}\! +\! {\begin{bmatrix}
0\\
1
\end{bmatrix}}\left( {u\left( t \right)\! +\! {\Theta ^{\rm{T}}}\left( t \right)\Phi \left( {x\left( t \right)} \right)} \right),	  
\end{equation}
where ${x_1}\left( t \right)$ is the roll angle, ${x_2}\left( t \right)$ is the roll rate, $u\left( t \right)$ is the virtual control, $\Phi \left( x \right) = {{\begin{bmatrix}
{{x_1}}&{{x_2}}&{\left| {{x_1}} \right|{x_2}}&{\left| {{x_2}} \right|{x_2}}&{x_1^3}
\end{bmatrix}}^{\rm{T}}}$. Similar to the experiments in \cite{b22}, the numerical value of the parameters $\Theta \left( t \right)$ was taken from \cite{b31} and increased by a factor of 1000. The jump change of the uncertainty parameters simulated an instantaneous, comparing to the wing rock dynamics, change in the attack angle from 15 to 25 degrees and back. Therefore, the parameters in \eqref{eq4} were defined as:
\begin{equation}\label{eq31}
\begin{array}{l}
{\Theta _0} = {{\begin{bmatrix}
{3.63}&{ - 8.58}&{20.2}&{ - 21.9}&{ - 51.88}
\end{bmatrix}}^{\rm{T}}}{\rm{;}}\;\\
{\theta _1} = {{\begin{bmatrix}
{ - 22.22}&{23.74}&{ - 82.66}&{31.45}&{73.33}
\end{bmatrix}}^{\rm{T}}}{\rm{;}}\;\\
{\theta _2} =  - {\theta _1}{\rm{;}}\quad j = \left\{ {1{\rm{;\;2}}} \right\}.
\end{array}
\end{equation}

The reference signal \eqref{eq6} was implemented as a square wave:
\begin{equation}\label{eq32}
r\left( t \right) = \left\{ \begin{array}{l}
1,{\rm{\;for\;}}0{\rm{ < }}t \le 8,\\
0,{\rm{\;for\;}}8{\rm{ < }}t \le 16,\\
1,{\rm{\;for\;}}16{\rm{ < }}t \le 24.
\end{array} \right.
\end{equation}

To demonstrate the performance of the system in all three cases considered in the stability analysis, at ${t_1} = 4$ seconds the parameter ${\Theta _0}$ was changed by the value of ${\theta _1}$, and at ${t_2} = 17$ the sum of ${\Theta _0} + {\theta _1}$ was changed by the value of ${\theta _2}$. The parameters of the baseline controller ${u_{bl}}\left( t \right)$ were calculated according to the method of the LQ synthesis by optimization of the following quality criterion:
\begin{equation}\label{eq33}
J = \int\limits_0^\infty  {\left( {{x^{\rm{T}}}\left( \tau  \right){Q_{LQ}}x\left( \tau  \right) + u_{bl}^{\rm{T}}\left( \tau  \right){R_{LQ}}{u_{bl}}\left( \tau  \right)} \right)} {\rm{}}\;d\tau {\rm{,}}
\end{equation}
where ${Q_{LQ}} = diag\left\{ {2800,{\rm{\;1}}} \right\}$, ${R_{LQ}} = 100$. The values of other constants of the adaptive control system are shown in Table I.

The aim of the experiment was to compare the developed system (CMRAC) with the solution based on the conventional adaptive law ${\Gamma _1}\Phi \left( x \right)e_{ref}^{\rm{T}}PB$ (MRAC) when the same value of the adaptation gain ${\Gamma _1} = 500{I_{5 \times 5}}$ was used in both laws. 
\begin{table}\label{table1}
\centering
\caption{Simulation Parameters}
\setlength{\tabcolsep}{3pt}
\begin{tabular}{|c|c|c|c|}
\hline
Parameter& 
Value& 
Parameter&
Value\\
\hline
{$\hat \Theta \left( 0 \right)$}&	$0_{5 \times 1}$&	$\Gamma_2 \left(0\right)$&	1\\
$Q$&    $diag\left\{100,10\right\}$& $\lambda_2$&    450\\
$PB$&   $\left[9.45;\;4.43\right]$& $l$&    10\\
$\Gamma_1$& $500I_{5 \times 5}$&    $k$&    10\\
$\lambda_1$&    1100&   $\sigma$&   5\\
\hline
\end{tabular}
\end{table}

Fig.1 shows the change of the reset time $t_r^ + $ value in the course of the experiment.
\vspace{-12pt}
\begin{figure}[thpb]\label{figure1}
\begin{center}
\includegraphics[scale=0.5]{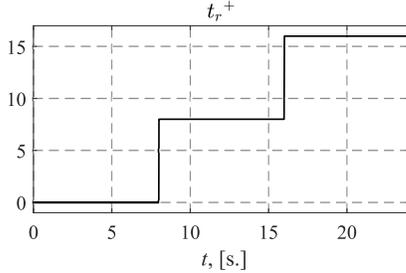}
\caption{Reset time $t_r^ +$  curve in the course of the experiment.}
\end{center}
\end{figure}
\vspace{-14pt}

It follows from Fig. 1 that the reset of the filtering procedure \eqref{eq11}, \eqref{eq12}, \eqref{eq14}, \eqref{eq17} was made strictly at the moments of $r\left( t \right)$ change.

Fig. 2 shows the transients of the regressor $\Omega \left( t \right)$ at time intervals corresponded to change either of $r\left( t \right)$ or the uncertainty parameters.
\vspace{-12pt}
\begin{figure}[thpb]\label{figure2}
\begin{center}
\includegraphics[scale=0.6]{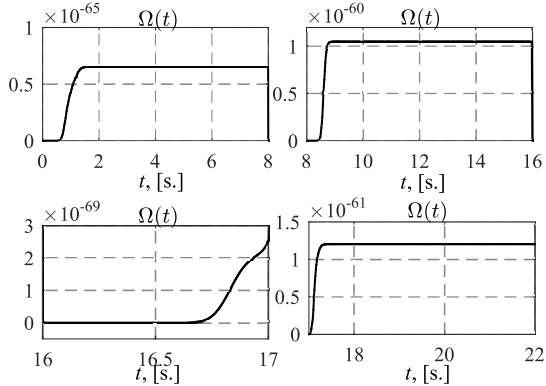}
\caption{Transients of the regressor $\Omega \left( t \right)$.} 
\end{center}
\end{figure}
\setlength{\textfloatsep}{0pt plus 1.0pt minus 2.0pt}
\vspace{-14pt}

Considering the regressor $\Omega \left( t \right)$, its transients in Fig. 2 proved the conclusions about its properties drawn in Proposition \ref{proposition3}. The low magnitude $\left( {{\rm{\sim 1}}{{\rm{0}}^{ - 61}} \div {\rm{1}}{{\rm{0}}^{ - 69}}} \right)$ of $\Omega \left( t \right)$ is explained by the fact that the multiplication by the adjoint matrix is used in \eqref{eq15} . The difference of about $\left( {{\rm{\sim 1}}{{\rm{0}}^8}} \right)$ between $\Omega \left( t \right)$ values at different stages of the experiment verified the need to adjust ${\Gamma _2}\left( t \right)$.

The transients of the control action $u\left( t \right)$ (a), states $x\left( t \right)$ (b), and the estimates of the unknown parameters (c) are shown in Fig. 3.

So, for the developed system, the oscillations of curves of the control signal $u\left( t \right)$ and state vector $x\left( t \right)$ had existed exactly till the moment when the uncertainty ideal parameters were identified. Using the proposed law \eqref{eq18}, the parameter error was always bounded, and exponentially stable over the intervals $\left[ {0{\rm{;4}}} \right]$ and $\left[ {8{\rm{;16}}} \right]$. This verified the proof of Theorems 1-3 and the conclusions made in Remark \ref{remark3}.

\begin{figure}[!t]\label{figure3}
\begin{center}
\includegraphics[scale=0.40]{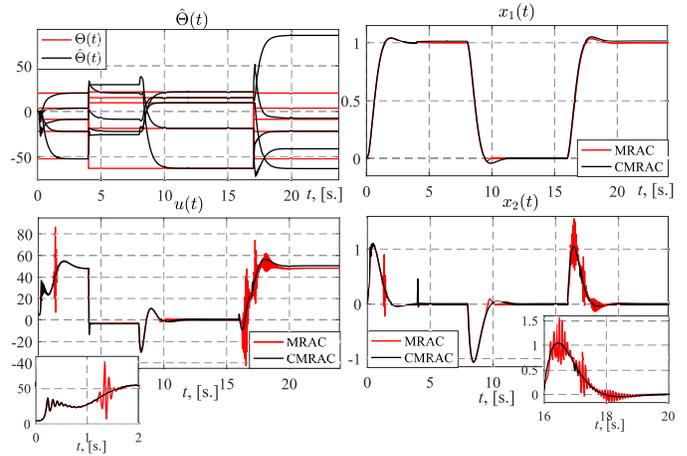}
\caption{Transients of unknown parameters estimates $\hat \Theta \left( t \right)$, control action $u\left( t \right)$ and states $x\left( t \right)$.} 
\end{center}
\end{figure}

Figure 4 shows the transients for the state ${x_1}\left( t \right)$ and the augmented tracking error $\xi \left( t \right)$ obtained by application of the developed adaptive law \eqref{eq18}, the classical law ${\Gamma _1}\Phi \left( x \right)e_{ref}^{\rm{T}}PB$, and the composite laws \eqref{eq23}, \eqref{eq27}, \eqref{eq29} with \eqref{eq28}, and \eqref{eq29} with \eqref{eq29i5}.

\begin{figure}[!t]\label{figure4}
\begin{center}
\includegraphics[scale=0.37]{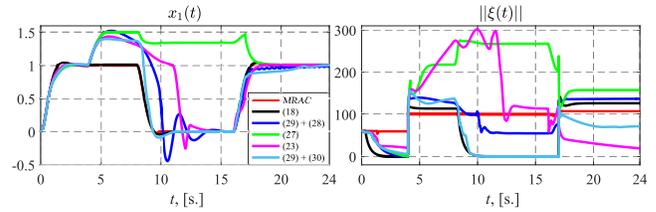}
\caption{{Transients of state $x_1$ (a) and augmented error (b).}} 
\end{center}
\end{figure}

Table II contains the parameters values of the laws \eqref{eq23}, \eqref{eq27}, \eqref{eq29} with \eqref{eq28}, and \eqref{eq29} with \eqref{eq29i5}, which were used for the simulation. The values of all other parameters were set according to Table I.

\begin{table}\label{table2}
\centering
\caption{Simulation Parameters}
\setlength{\tabcolsep}{3pt}
\begin{tabular}{|c|c|c|c|}
\hline
Parameter& 
Value& 
Parameter&
Value\\
\hline
$k_l$&	5&	$l$ for \eqref{eq28}&   10\\
$k_{ll}$&	1000&	$\lambda _{\min }^{LB}$&   $10^{-12}$\\
$k_{sw}$&	50000&	$\lambda _{\min }^{UB}$&   $10^{-5}$\\
$l_m$&	0.1&	$\Gamma_2$ for \eqref{eq27}&   2500\\
$l_M = l_0$&	10&	$\Gamma_2$ for \eqref{eq29} with \eqref{eq28}&   2500\\
$\vartheta$&    1& $\Gamma_2$ for \eqref{eq29} with \eqref{eq29i5}&   1000\\
\hline
\end{tabular}
\end{table}

{As follows from the transients presented in Fig. 4, over the interval $\left[ {0{\rm{; 4}}} \right]$ all composite laws \eqref{eq18}, \eqref{eq23}, \eqref{eq27}, \eqref{eq29} with \eqref{eq28} and \eqref{eq29} with \eqref{eq29i5} ensured exponential convergence of the augmented error to zero. At the same time, when the uncertainty parameters had changed at time instant $t=4$, only the adaptive laws \eqref{eq18} and \eqref{eq29} with \eqref{eq29i5} preserved the property of exponential convergence of the augmented tracking error to zero after the reference value change at time instant} \linebreak { $t=8$. Here, it should be specially noted that i) the adaptive law \eqref{eq29} with \eqref{eq29i5} ensured $\xi \left( t \right) \in {\rm{GES}}$ for $t \ge 8$ because ${\lambda _{\min }}\left( {\varphi \left( t \right)} \right)$ kept the value, which was close to $\lambda _{\min }^{UB}$, over a sufficiently long time range; ii) the adaptive law \eqref{eq18} ensured $\xi \left( t \right) \in {\rm{GES}}$ for} \linebreak {$t \ge 8$ because uncertainty parameters had changed their values before change of the reference value; iii) considering the time interval $t \ge {t_2} = 17$, the exponential stability of the augmented error $\xi \left( t \right)$ was provided by \eqref{eq23} owing to the fact that ${y_{sw}}\left( t \right)$ stored the data about the true regression parameters ${\Theta _0}$, which were recorded to ${y_{sw}}\left( t \right)$ over the interval $\left[ {0{\rm{; 4}}} \right]$ according to \eqref{eq24} at time point $T$.}

Also, the transients shown in Fig.4 demonstrated the advantages of the proposed system of composite adaptive control over the existing ones. Only the developed system provides strict guarantees to avoid the superpositional mixing of the data on the regressions with different parameters after each $r\left(t\right)$ value change (if the uncertainty parameters had changed their values before that). Fig. 4 also clearly shows that the control quality, which was provided by the known composite control systems when the superpositional mixing happened, was sufficiently worse than the one by the conventional MRAC. Under such conditions, the law \eqref{eq18} deteriorated the transients quality insignificantly. Some extended simulation results are shown in Supplementary Material \cite{b24}.

In general, the results of the experiments validated the analytically proved property of the developed system to guarantee exponential convergence of the parameter error when the FE requirement of the regressor was met and the uncertainty parameters changed before the change of the reference $r\left( t \right)$. Also, in comparison with the conventional law, it was possible to improve the quality of the transients of ${e_{ref}}\left( t \right)$ and $u\left( t \right)$ after the completion of the uncertainty parameters identification.

\section{Conclusion and Future Work}
In this paper, in order to relax PE requirement for the MRAC scheme with the piecewise-constant uncertainty parameters of the plant, the method was proposed, which was {based on a novel scheme of uncertainty filtration with resetting.} Such scheme made it possible to develop a CMRAC adaptation law, which guaranteed exponential convergence of the parameter error to zero if FE requirement was met and the following conditions were satisfied: 1) the reference $r\left( t \right)$ was a piecewise-constant signal; 2) a change of $r\left( t \right)$ caused the regressor finite excitation; and 3) the unknown piecewise-constant parameters had already changed their values before the change of $r\left( t \right)$. The analytical stability analysis, as well as the conducted numerical experiments, demonstrated the main properties of the obtained system and verified the paper contribution. The proposed method differs from the existing ones, which are also used to relax PE requirement for MRAC, by application of the filters \eqref{eq11}, \eqref{eq12}, \eqref{eq14} and \eqref{eq16}  with resetting. It relaxes PE requirement not only for the constant unknown uncertainty parameters, but also for the piecewise-constant ones under some weak additional assumptions.

In further research, we plan: 1) to improve the transient response of the obtained system over the intervals when the ideal uncertainty parameters have not been found yet; {2) to extend the obtained results to Case 2 and 3 by development of a robust resetting scheme based on an algorithm to detect uncertainty parameters change (some preliminary results in this sense can be found in \cite{b32},{\cite{b33}}).}


\begin{thebibliography}{00}
\bibitem{b1} I. D. Landau, ``A survey of model reference adaptive techniques—Theory and applications,'' \emph{Automatica}, vol. 10, no. 4, pp. 353--379, 1974.
\bibitem{b2} R. Kumar, et al. ``Review on model reference adaptive system for sensorless vector control of induction motor drives,'' \emph{IET Electric Power Applications,} vol. 9, no. 7, pp. 496--511, 2015.
\bibitem{b3} M. Korzonek, G. Tarchala, and T. Orlowska-Kowalska, ``A review on MRAS-type speed estimators for reliable and efficient induction motor drives,” \emph{ISA transactions,} vol. 93, pp. 1--13, 2019.
\bibitem{b4} K. S. Narendra, and L. S. Valavani, ``Direct and indirect model reference adaptive control,'' \emph{Automatica,} vol. 15, no. 6, pp. 653--664, 1979.
\bibitem{b5} E. Lavretsky, ``Combined/Composite Model Reference Adaptive Control,'' \emph{IEEE Trans. on Automatic Control,} vol. 54, pp.2692–-2697, 2009.
\bibitem{b6} K. S. Narendra, and A. M. Annaswamy, \emph{Stable adaptive systems,} Mineola, NY, USA: Courier Corp., 2005.
\bibitem{b7} K. S. Narendra, and A. M. Annaswamy, ``Persistent excitation in adaptive systems,'' \emph{Int. Journal of Control,} vol. 45, no. 1, pp. 127--160, 1987.
\bibitem{b8} B. M. Jenkins, A. M. Annaswamy, E. Lavretsky, and T. E. Gibson, ``Convergence properties of adaptive systems and the definition of exponential stability,'' \emph{SIAM Journal on Control and Optimization,} vol. 56, no. 4, pp. 2463--2484, 2018.
\bibitem{b9} P. A. Ioannou, and J. Sun, \emph{Robust adaptive control,} Mineola, NY, USA: Courier Corp., 2012.
\bibitem{b10} S. Boyd, and S. S. Sastry, ``Necessary and sufficient conditions for parameter convergence in adaptive control,'' \emph{Automatica,} vol. 22, no. 6, pp. 629--639, 1986.
\bibitem{b11} R. Ortega, V. Nikiforov, and D. Gerasimov, ``On modified parameter estimators for identification and adaptive control. A unified framework and some new schemes,'' \emph{Annual Reviews in Control,} pp.1--16, 2020.
\bibitem{b12} L. Ljung, and T. Söderström, \emph{Theory and practice of recursive identification,} MIT press series in signal proc., opt., and control, 1983.
\bibitem{b13} E. N. Johnson, and S. M. Oh, ``Adaptive control using combined online and background learning neural network,'' in \emph{Proc. IEEE Conference on Decision and Control (CDC),} vol. 5, Bahamas, 2004, pp. 5433--5438.
\bibitem{b14} A. Kutay, G. Chowdhary, A. Calise, and E. Johnson, ``A comparison of select direct adaptive control methods under actuator failure accommodation,'' in \emph{Proc. AIAA Guidance, Navigation and Control Conference and Exhibit.,} AIAA, Reston, VA, 2008.
\bibitem{b15} G. Chowdhary, and E. Johnson, ``Theory and flight test validation of long term learning adaptive flight controller,'' in \emph{Proc. AIAA Guidance, Navigation and Control Conf. and Exhibit,} AIAA, Reston, VA, 2008.
\bibitem{b16} G. Chowdhary, T. Yucelen, M. Muhlegg, and E. Johnson, ``Concurrent learning adaptive control of linear systems with exponentially convergent bounds,'' \emph{International Journal of Adaptive Control and Signal Processing,} vol. 27, no. 4, pp. 280--301, 2013.
\bibitem{b17} G. Chowdhary, M. Mühlegg, and E. Johnson, ``Exponential parameter and tracking error convergence guarantees for adaptive controllers without persistency of excitation,'' \emph{International Journal of Control,} vol. 87, no. 8, pp. 1583--1603, 2014.
\bibitem{b18} S. B. Roy, S. Bhasin, and I. N. Kar, ``Combined MRAC for unknown MIMO LTI systems with parameter convergence,'' \emph{IEEE Transactions on Automatic Control,} vol. 63, no. 1, pp. 283--290, 2017.
\bibitem{b19} G. Kreisselmeier, ``Adaptive observers with exponential rate of convergence,'' \emph{IEEE Trans. on Automatic Control,} vol.22, no. 1, pp. 2--8, 1977.
\bibitem{b20} A. Glushchenko, V. Petrov, and K. Lastochkin ``Robust method to provide exponential convergence of model parameters solving linear time-invariant plant identification problem,'' \emph{Int. Journal of Adaptive Control and Signal Processing,} vol. 35, no. 6, pp. 1120--1137, 2021.
\bibitem{b21} H. I. Lee, H. S. Shin, A. Tsourdos, ``Concurrent learning adaptive control with directional forgetting,'' \emph{IEEE Trans. on Autom. Control,} vol. 64, no. 12, pp. 5164--5170, 2019.
\bibitem{b22} N. Cho, H. S. Shin, Y. Kim, and A. Tsourdos, ``Composite model reference adaptive control with parameter convergence under finite excitation,'' \emph{IEEE Trans. on Autom. Control,} vol.63, no.3, pp.811--818, 2017.
{\bibitem{b22i5} Y. Pan, S. Aranovskiy, A. Bobtsov, H. Yu, ``Efficient learning from adaptive control under sufficient excitation,'' \emph{International Journal of Robust and Nonlinear Control}, vol. 29, no. 10, pp. 3111--3124, 2019.}
\bibitem{b23} G. Chowdhary, M. Mühlegg, J. How, and F. Holzapfel ``A concurrent learning adaptive-optimal control architecture for nonlinear systems,'' in \emph{Proc. IEEE Conf. on Decision and Control,} Florence, 2013, pp. 868--873.
\bibitem{b25} S. Aranovskiy, A. Bobtsov, R. Ortega, and A. Pyrkin, ``Performance enhancement of parameter estimators via dynamic regressor extension and mixing,'' \emph{IEEE Transactions on Automatic Control,}  vol. 62, no. 7, pp. 3546--3550, 2016.
\bibitem{b24} A. Glushchenko, V. Petrov, and K. Lastochkin, ``Supplement to 'Regression Filtration with Resetting to Provide Exponential Convergence of MRAC for Plants with Jump Change of Unknown Parameters' in IEEE-TAC'', 2022. \emph{Preprint}. Available: \newline {\scriptsize \url{https://arxiv.org/src/2102.10359v4/anc/supp.pdf}}.
\bibitem{b26} S. Aranovskiy, A. Belov, R. Ortega, N. Barabanov, and A. Bobtsov, ``Parameter identification of linear time‐invariant systems using dynamic regressor extension and mixing,'' \emph{International Journal of Adaptive Control and Signal Processing,} vol. 33, no. 6, pp. 1016--1030, 2019.
\bibitem{b27} S. Aranovskiy, R. Ushirobira, M. Korotina, and A. Vedyakov ``On preserving-excitation properties of a dynamic regressor extension scheme,'' {\emph{IEEE Transactions on Automatic Control,} pp. 1-6, 2022}.
\bibitem{b28} M. Korotina, S. Aranovskiy, R. Ushirobira, and A. Vedyakov, ``On parameter tuning and convergence properties of the DREM procedure,'' in \emph{Proc. Europ. Control Conf.,} Saint-Petersburg, Russia, 2020, pp. 1--7.
\bibitem{b29} A. Glushchenko, V. Petrov, and K. Lastochkin, ``I-DREM: Relaxing the Square Integrability Condition,'' \emph{Automation and Remote Control,} vol. 82, no. 7, pp. 1233--1247, 2021.
\bibitem{b30}	S. B. Roy, S. Bhasin, and I. N. Kar, ``A UGES switched MRAC architecture using initial excitation,'' \emph{IFAC-PapersOnLine,} vol. 50, no. 1, pp. 7044--7051, 2017.
\bibitem{b31} J. M. Elzebda, A. H. Nayfeh, and D. T. Mook, ``Development of an Analytical Model of Wing Rock for Slender Delta Wings,” \emph{Journal of Aircraft,} vol. 26, no. 8, pp. 737–-743, 1989.
{\bibitem{b32} A. Glushchenko, and K. Lastochkin, ``Unknown Piecewise Constant Parameters Identification with Exponential Rate of Convergence,'' 2022. \emph{Preprint.} Available: {\scriptsize \url{https://arxiv.org/pdf/2203.11685.pdf}}.}
{\bibitem{b33}{A. Glushchenko, and K. Lastochkin, ``Exponentially Stable MRAC of MIMO Switched Systems with Matched Uncertainty and Completely Unknown Control Matrix,'' 2022. \emph{Preprint.} Available:\linebreak {\scriptsize \url{https://arxiv.org/pdf/2208.03972.pdf}}.}}
\end{thebibliography}
\end{document}